\documentclass[
  aip,
  jcp,
  reprint,
  amsmath,
  amssymb,
  floatfix
]{revtex4-2}

\usepackage{mathtools}
\usepackage{bm}
\usepackage{braket}

\usepackage[version=4]{mhchem}
\usepackage{graphicx}

\begin{document}

\title{Importance-Reweighted Fock-Space Variational Monte Carlo}

\author{Zheng Che}
\email{wsmxcz@gmail.com}
\affiliation{Hefei National Research Center for Physical Sciences at the Microscale, University of Science and Technology of China, Hefei 230026, China}

\begin{abstract}
Fock-space variational Monte Carlo (FS-VMC) evaluates variational quantities by sampling discrete many-body configurations. In molecular applications, concentrated Born distributions can hinder Markov-chain mixing, while Monte Carlo estimators can also exhibit large variance. We introduce importance-reweighted FS-VMC (IR-FS-VMC), which combines Metropolis--Hastings sampling, an evaluation kernel, and self-normalized importance sampling to estimate the Born-distribution energy, gradient, and stochastic reconfiguration (SR) matrix. Controlled comparisons on equilibrium \ce{H2O} and \ce{Fe2S2} show that the acceptance rate and local-energy variance can respond differently to the sampling and estimation procedure. A single production protocol yields accurate variational energies for \ce{H2O} dissociation, 36-site hydrogen lattices, and \ce{Fe2S2} and \ce{Fe4S4} active spaces without system-specific sampling parameters.
\end{abstract}

\maketitle

\section{Introduction}

Accurate low-energy solutions of interacting electronic Hamiltonians remain a central problem in quantum chemistry and quantum many-body physics. In a finite orbital basis and fixed symmetry sector, the many-body state is represented in a Fock space whose dimension grows combinatorially with system size. Full configuration interaction is exact within this space but scales exponentially, motivating complementary approximations such as coupled-cluster theory,\cite{bartlett2007coupled} selected configuration interaction,\cite{holmes2016heatbath,tubman2016deterministic} and matrix-product-state density matrix renormalization group (DMRG).\cite{chan2002dmrg,chan2011density,sharma2012spinadapted}

Variational Monte Carlo (VMC) provides a complementary route by optimizing an explicit parametrized wave function from stochastic estimates of the Rayleigh quotient and its derivatives.\cite{foulkes2001quantum} In a discrete many-body basis, Fock-space VMC (FS-VMC) evaluates these quantities from sampled configurations and sparse Hamiltonian connections. Related formulations have long been used for lattice Hamiltonians,\cite{yokoyama1987variational,tahara2008variational,sandvik2007variational,nomura2017restricted,choo2019j1j2,viteritti2023transformer} and have also been developed for molecular electronic structure.\cite{neuscamman2012optimizing,wei2018reduced,sabzevari2018orbital,rath2023framework} Neural quantum states have substantially enlarged the available variational families,\cite{carleo2017solving,choo2020fermionic,zhao2023scalable,luo2019backflow,liu2024backflow,wu2025hybrid,li2026spinadapted} while stochastic reconfiguration and related natural-gradient methods enable the optimization of increasingly flexible parametrizations.\cite{sorella2001generalized,chen2024minsr,peng2025optimization}

Electronic-structure problems span widely different correlation regimes and many-body structures, posing a general challenge for efficient sampling and low-variance estimation across systems and geometries. Selected-configuration approaches avoid stochastic sampling by exploiting sparsity of the wave function in configuration space,\cite{li2023nonstochastic,che2026deterministic,solanki2026selected} while autoregressive wave functions permit direct Born-distribution sampling without Markov-chain autocorrelation.\cite{barrett2022autoregressive} The former relies on a sparsity assumption, whereas the latter requires an autoregressive factorization that is not generally available to fermionic backflow ansatzes.\cite{luo2019backflow,liu2024backflow} Direct sampling removes Markov-chain autocorrelation but does not guarantee lower variance.

Importance sampling provides a more general way to modify the statistical distribution used for Monte Carlo estimation while retaining the original variational quantities.\cite{trail2008alternative,trail2010optimum,misery2026importance} In Fock space, sparse Hamiltonian connections provide a natural structure for constructing Metropolis--Hastings proposal kernels,\cite{holmes2016efficient} while stochastic post-sampling kernels can further modify the distribution entering local estimators.\cite{wan2026removing} These ingredients allow sampling and estimation to be adjusted without imposing additional structure on the wave-function ansatz.

Here we develop importance-reweighted Fock-space variational Monte Carlo (IR-FS-VMC). A Metropolis--Hastings chain samples an invariant distribution, which is mapped by an evaluation kernel to the importance distribution. We first compare Born, tempered, and IR variants for equilibrium \ce{H2O} and \ce{Fe2S2}, and then apply a single production protocol to \ce{H2O} bond breaking, 36-site hydrogen lattices, and \ce{Fe2S2} and \ce{Fe4S4} active spaces, spanning qualitatively different correlation regimes.

\section{Theory}
\label{sec:theory}

\subsection{Fock-Space Variational Monte Carlo}
\label{sec:variational_formulation}

Let \(\mathcal X\) be a finite configuration space in a fixed Fock-space sector, with orthonormal basis states \(\{\ket{x}:x\in\mathcal X\}\). For a real parametrized wave function
\begin{equation}
\ket{\psi_{\theta}}=\sum_{x\in\mathcal X}\psi_{\theta}(x)\ket{x},
\label{eq:variational_state}
\end{equation}
the variational energy is
\begin{equation}
E(\theta)=
\frac{\braket{\psi_{\theta}|\hat H|\psi_{\theta}}}
{\braket{\psi_{\theta}|\psi_{\theta}}}.
\label{eq:rayleigh}
\end{equation}
Define the Born distribution
\begin{equation}
p_{\theta}(x)=\frac{|\psi_{\theta}(x)|^2}{Z_{\theta}},
\qquad
Z_{\theta}=\sum_x|\psi_{\theta}(x)|^2,
\label{eq:born}
\end{equation}
and, where \(\psi_{\theta}(x)\neq0\), the local energy
\begin{equation}
E_{\mathrm L}(x)=
\frac{\braket{x|\hat H|\psi_{\theta}}}{\psi_{\theta}(x)}
=\sum_y H_{xy}\frac{\psi_{\theta}(y)}{\psi_{\theta}(x)},
\label{eq:local_energy}
\end{equation}
with \(H_{xy}=\braket{x|\hat H|y}\). The energy is then \(E=\mathbb E_{p_{\theta}}[E_{\mathrm L}]\).

For the logarithmic derivatives
\begin{equation}
O_i(x)=\partial_{\theta_i}\log|\psi_{\theta}(x)|,
\label{eq:log_derivative}
\end{equation}
the energy gradient and SR matrix are
\begin{align}
g_i&=2\mathbb E_{p_{\theta}}[\Delta O_i\,\Delta E_{\mathrm L}],
\label{eq:energy_gradient}\\
S_{ij}&=\mathbb E_{p_{\theta}}[\Delta O_i\,\Delta O_j],
\label{eq:metric}
\end{align}
where \(\Delta A=A-\mathbb E_{p_{\theta}}[A]\). The matrix \(S\) is the real part of the quantum geometric tensor used in SR.\cite{sorella2001generalized,stokes2020quantum} The energy, gradient, and SR matrix are therefore Born-distribution expectations.

\begin{figure}[t]
  \centering
  \includegraphics[width=\columnwidth]{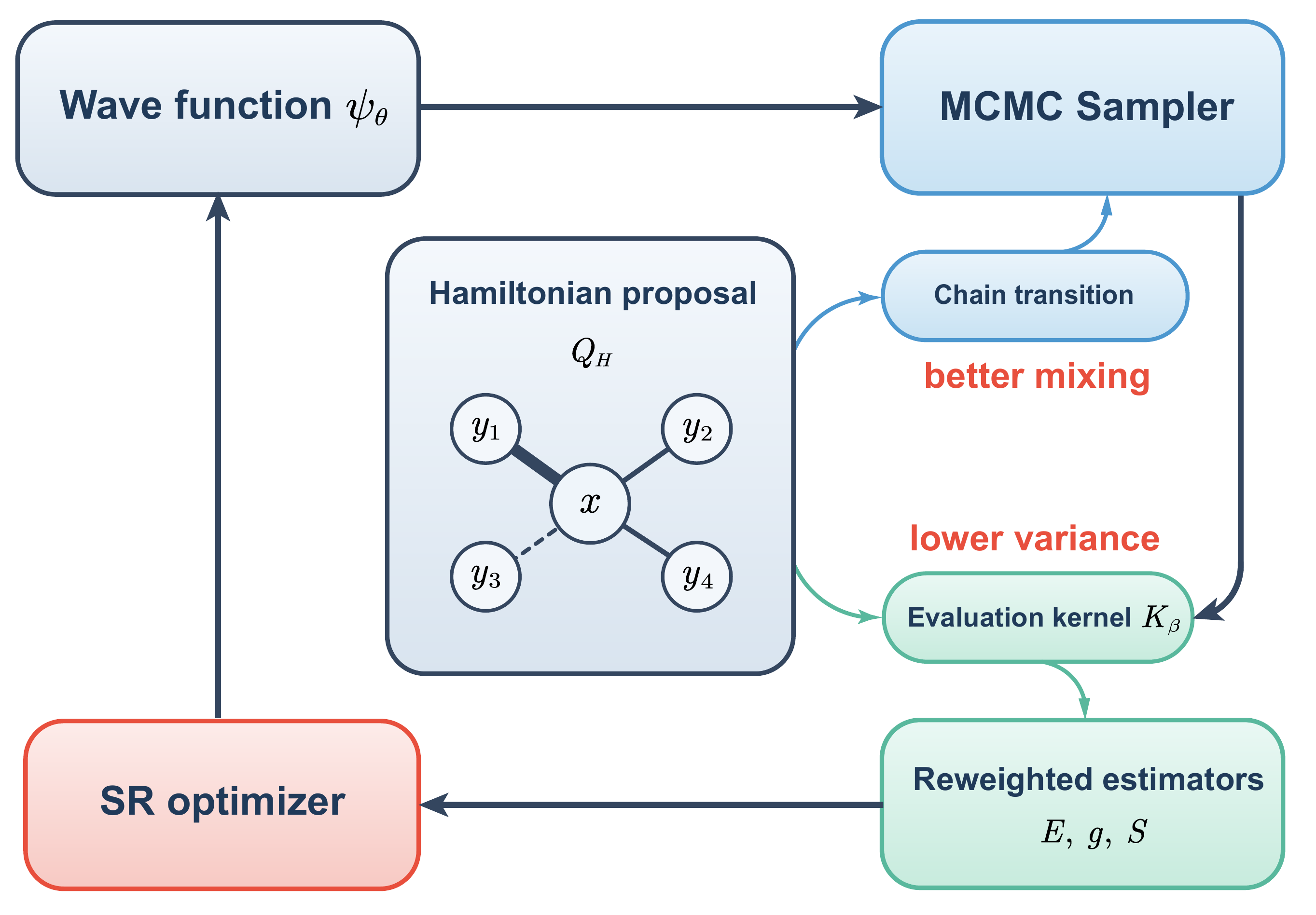}
  \caption{Overview of IR-FS-VMC. Metropolis--Hastings sampling uses the proposal kernel $Q_H$, and the evaluation kernel $K_\beta$ defines the configurations entering self-normalized importance estimators of Born-distribution quantities.}
  \label{fig:overview}
\end{figure}

\subsection{Importance Reweighting}
\label{sec:reweighted_estimators}

Let $r_{\theta}(x)=\widetilde r_{\theta}(x)/Z_r$ be an importance distribution satisfying $\operatorname{supp}(p_{\theta})\subseteq\operatorname{supp}(r_{\theta})$. Define the unnormalized importance weight
\begin{equation}
\omega_{\theta}(x)=
\frac{|\psi_{\theta}(x)|^2}{\widetilde r_{\theta}(x)}.
\label{eq:importance_weight}
\end{equation}
For any integrable quantity $f$,
\begin{equation}
\mathbb E_{p_{\theta}}[f]
=
\frac{\mathbb E_{r_{\theta}}[\omega_{\theta}f]}
{\mathbb E_{r_{\theta}}[\omega_{\theta}]},
\label{eq:reweighting_identity}
\end{equation}
so the unknown normalization constants cancel.\cite{trail2008alternative,trail2010optimum}

For a stationary sequence $\{x_n\}_{n=1}^{N_{\mathrm s}}$ with marginal distribution $r_{\theta}$, define
\begin{equation}
w_n=\frac{\omega_{\theta}(x_n)}{\sum_m\omega_{\theta}(x_m)},
\qquad
\overline A=\sum_n w_n A(x_n).
\label{eq:normalized_weights}
\end{equation}
The corresponding self-normalized estimators are
\begin{align}
\widehat E&=\overline E_{\mathrm L},
\label{eq:reweighted_energy}\\
\widehat g_i&=2\sum_n w_n
[O_i(x_n)-\overline O_i]
[E_{\mathrm L}(x_n)-\overline E_{\mathrm L}],
\label{eq:reweighted_gradient}\\
\widehat S_{ij}&=\sum_n w_n
[O_i(x_n)-\overline O_i]
[O_j(x_n)-\overline O_j].
\label{eq:reweighted_metric}
\end{align}
Self-normalization introduces finite-sample bias, but the estimators remain consistent under the usual ergodicity and moment conditions. The same normalized weights apply to other Born-distribution expectations. For correlated samples, the asymptotic variance depends on the autocovariance sequence of the importance-weighted integrand; details are given in the supplementary material.

\subsection{Sampling and Estimation}
\label{sec:sampling_estimation}

For the ab initio calculations, we use the nonrelativistic electronic Hamiltonian in an orthonormal spatial-orbital basis,
\begin{equation}
\hat H=E_{\mathrm{core}}
+\sum_{pq\sigma}h_{pq}a_{p\sigma}^{\dagger}a_{q\sigma}
+\frac12\sum_{pqrs\sigma\tau}(pq|rs)
a_{p\sigma}^{\dagger}a_{r\tau}^{\dagger}a_{s\tau}a_{q\sigma}.
\label{eq:electronic_hamiltonian}
\end{equation}
Let $\mathcal C(x)$ denote a symmetric retained set of nonzero off-diagonal Hamiltonian connections, $y\in\mathcal C(x)\Leftrightarrow x\in\mathcal C(y)$, and define the normalization factor
\begin{equation}
d(x)=\sum_{y\in\mathcal C(x)}|H_{xy}|.
\label{eq:proposal_normalization}
\end{equation}
The Metropolis--Hastings proposal kernel is
\begin{equation}
Q_H(y|x)=\frac{|H_{xy}|}{d(x)},
\qquad y\in\mathcal C(x),
\label{eq:proposal_kernel}
\end{equation}
so that proposal probabilities are proportional to $|H_{xy}|$ over the retained connections.\cite{holmes2016efficient}
We choose the invariant distribution
\begin{equation}
\rho_{\theta,\alpha}(x)\propto
d(x)|\psi_{\theta}(x)|^{\alpha},
\qquad 0\le\alpha\le2,
\label{eq:invariant_distribution}
\end{equation}
for which $d(x)$ cancels from the Metropolis--Hastings ratio,
\begin{equation}
a(x,y)=
\min\left\{1,
\left|\frac{\psi_{\theta}(y)}{\psi_{\theta}(x)}\right|^{\alpha}
\right\}.
\label{eq:acceptance_probability}
\end{equation}
The exponent $\alpha$ controls the amplitude dependence of the invariant distribution and is adapted during optimization as described in the supplementary material.

The configurations entering the estimators are generated from
\begin{equation}
K_{\beta}(y|x)=
(1-\beta)\delta_{xy}+\beta Q_H(y|x),
\qquad 0\le\beta<1.
\label{eq:evaluation_kernel}
\end{equation}
The parameter $\beta$ selects between the current chain configuration and the proposed configuration for evaluation. Because this selection is independent of Metropolis--Hastings acceptance, rejected proposals can still contribute to the estimator. Marginalizing over the invariant distribution gives
\begin{equation}
\widetilde r_{\theta,\alpha,\beta}(y)=
(1-\beta)d(y)|\psi_{\theta}(y)|^{\alpha}
+\beta\sum_{x\in\mathcal C(y)}
|H_{xy}|\,|\psi_{\theta}(x)|^{\alpha}.
\label{eq:importance_distribution}
\end{equation}

Here $d(x)$ cancels between $\rho_{\theta,\alpha}(x)$ and $Q_H(y|x)$, leaving a single contraction over the Hamiltonian connections of $y$. This construction is related to blurred sampling, which also modifies the importance distribution through a stochastic kernel.\cite{wan2026removing} Applying $Q_H$ directly to Born samples would retain $1/d(x)$ for each incoming configuration and require the corresponding connection sums; the present choice cancels these factors analytically.

The importance weights in Eq.~(\ref{eq:importance_weight}) then recover the Born-distribution expectations in Eqs.~(\ref{eq:reweighted_energy})--(\ref{eq:reweighted_metric}).

\section{Computational Setup}
\label{sec:computational_setup}

For a two-body ab initio Hamiltonian, exact local-energy evaluation requires wave-function amplitudes on $O(N_{\mathrm{orb}}^4)$ single- and double-excitation connections in the worst case, where $N_{\mathrm{orb}}$ is the number of spatial orbitals. During optimization we therefore partition the off-diagonal Hamiltonian action into deterministic strong connections, $|H_{xy}|\ge\varepsilon_1$, stochastically contracted weak connections, $\varepsilon_2\le|H_{xy}|<\varepsilon_1$, and an omitted remainder below $\varepsilon_2$. The weak part is estimated from $N_{\mathrm{eloc}}$ stratified samples.\cite{petruzielo2012semistochastic,sharma2017semistochastic,wei2018reduced,sabzevari2018orbital,wu2025hybrid} This semistochastic contraction reduces the number of neural-wave-function evaluations.

The deterministic strong set is also used as the support of the Metropolis--Hastings proposal kernel, $\mathcal C(x)=\{y\ne x:|H_{xy}|\ge\varepsilon_1\}$, with integral-based screening.\cite{holmes2016efficient} Thus $\varepsilon_1$ fixes the retained connections used by both the local-energy contraction and the proposal kernel, whereas $\varepsilon_2$ and $N_{\mathrm{eloc}}$ affect only the local-energy estimator. For $\beta>0$, Eq.~(\ref{eq:importance_distribution}) requires one additional contraction over the retained connections.

All calculations use the same real spin-projected neural backflow ansatz,\cite{luo2019backflow,liu2024backflow} with configuration-dependent spatial orbitals, highest-weight spin projection, and reference-biased initialization. Parameters are optimized by sample-space stochastic reconfiguration\cite{sorella1998stochastic,sorella2001generalized,chen2024minsr} with a predictor based on the preceding update direction.\cite{goldshlager2024spring} The controlled comparisons keep the wave function, optimizer, sample size, and local-energy settings fixed and change only the sampling and estimation procedure.

The molecular and active-space Hamiltonians are expressed in orthonormal spatial orbitals. Published FCIDUMP Hamiltonians are used for Fe--S active spaces\cite{li2017spinprojected}, while \ce{H2O} and the 36-site hydrogen-lattice integrals are generated with PySCF.\cite{sun2018pyscf,sun2020pyscf} Independent spin-adapted DMRG reference calculations for the hydrogen lattices are performed with Block2.\cite{zhai2023block2,sharma2012spinadapted} The orbital bases, initial configurations, reference calculations, and numerical parameters are specified in the supplementary material.

All calculations were performed on a single NVIDIA A100 GPU (80 GB) with 16 Intel Xeon Platinum 8358 CPU cores and 128 GB of system memory.

\section{Results and Discussion}
\label{sec:results}

\subsection{Sampling and Optimization}
\label{sec:sampling_results}

We first compare five protocols for equilibrium \ce{H2O} and \ce{Fe2S2}. Born uses the Born distribution with a symmetric proposal kernel that selects single and double excitations with equal probability. Tempered uses an invariant distribution proportional to $|\psi|^\alpha$ with adaptive $\alpha$, the same proposal kernel, and reweighting to the Born distribution. The three IR variants use $(\alpha,\beta)=(2,0.5)$, adaptive $\alpha$ with $\beta=0$, and adaptive $\alpha$ with $\beta=0.5$; the last is the production protocol used elsewhere. The $\alpha=2$ variant omits amplitude tempering, while the $\beta=0$ variant has $K_0=I$ and evaluates local quantities at the chain configurations. All other settings are identical.

\begin{figure*}[t]
  \centering
  \includegraphics[width=\textwidth]{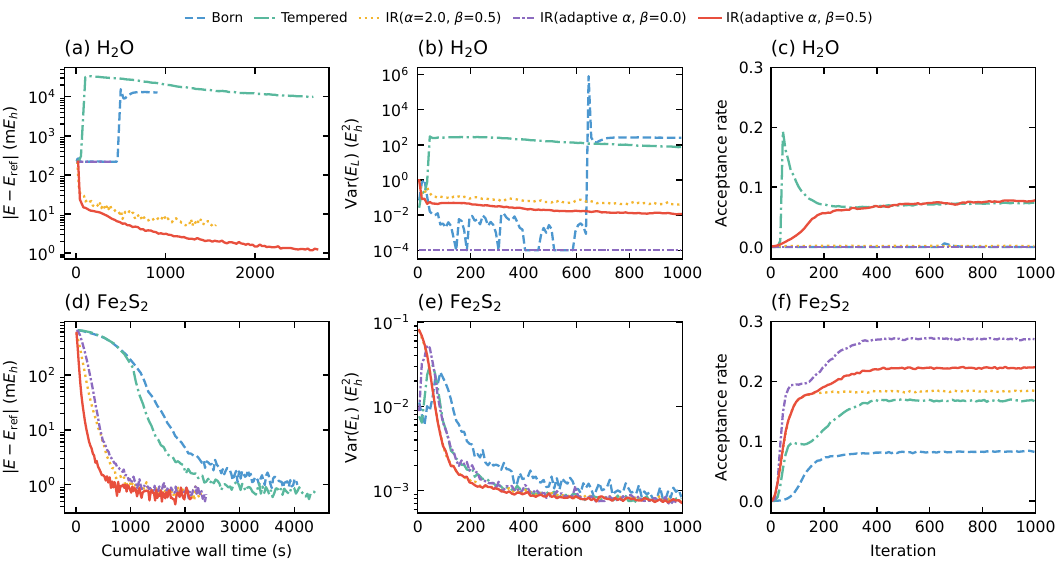}
  \caption{Sampling and optimization over the first $1000$ iterations for equilibrium \ce{H2O} (top) and \ce{Fe2S2} (bottom). Born uses the Born distribution with a symmetric proposal kernel that selects single and double excitations with equal probability; Tempered uses an invariant distribution proportional to $|\psi|^\alpha$ with adaptive $\alpha$ and the same proposal kernel; the three IR curves use $(\alpha,\beta)=(2,0.5)$, adaptive $\alpha$ with $\beta=0$, and adaptive $\alpha$ with $\beta=0.5$, respectively. Curves are medians over 10-iteration blocks. Left: $|E-E_{\mathrm{ref}}|$ versus cumulative wall time, with FCI and SA-DMRG references for \ce{H2O} and \ce{Fe2S2}. Center: sampled local-energy variance, with values below $10^{-4}\,E_{\mathrm h}^2$ shown at the plotting floor. Right: Metropolis--Hastings acceptance rate of the Markov chain.}
  \label{fig:sampling_optimization}
\end{figure*}

For equilibrium \ce{H2O} [Fig.~\ref{fig:sampling_optimization}(a)--(c)], the Born chain has near-zero acceptance and the optimization stalls. Over iterations $100$--$500$, Tempered raises the median acceptance rate to $0.069$, while the sampled local-energy variance reaches $2.51\times10^2\,E_{\mathrm h}^2$ and the energy remains far from FCI. In contrast, IR with $\alpha=2$ has a median acceptance rate of $0.0019$ and a much smaller local-energy variance of $8.03\times10^{-2}\,E_{\mathrm h}^2$, with a median energy error of $5.39\,\mathrm{m}E_{\mathrm h}$ over iterations $901$--$1000$.

The $\beta=0$ run remains close to its initial state, whereas the production IR protocol with adaptive $\alpha$ and $\beta=0.5$ reaches a median acceptance rate of $0.063$ and a local-energy variance of $3.15\times10^{-2}\,E_{\mathrm h}^2$ over iterations $100$--$500$, followed by a median error of $1.25\,\mathrm{m}E_{\mathrm h}$ over iterations $901$--$1000$.

\ce{Fe2S2} provides a contrasting regime [Fig.~\ref{fig:sampling_optimization}(d)--(f)]. The Born chain already has a median acceptance rate of $0.078$ over iterations $100$--$500$. The IR variants increase acceptance further, while all five protocols reduce the local-energy variance to the $10^{-3}\,E_{\mathrm h}^2$ scale and reach median errors of $0.63$--$1.07\,\mathrm{m}E_{\mathrm h}$ over iterations $901$--$1000$. Here the Born protocol is already effective, and the higher acceptance of the IR variants produces little change in the final energy trajectory. These comparisons show that increasing the acceptance rate does not necessarily reduce the local-energy variance or improve the optimization. Additional autocorrelation and effective-sample-size diagnostics are reported in the supplementary material.

\subsection{Variational Energies}
\label{sec:ground_state_energies}

We next assess the production protocol on systems spanning molecular bond breaking, extended hydrogen lattices, and transition-metal active spaces.

\begin{figure*}[t]
  \centering
  \includegraphics[width=\textwidth]{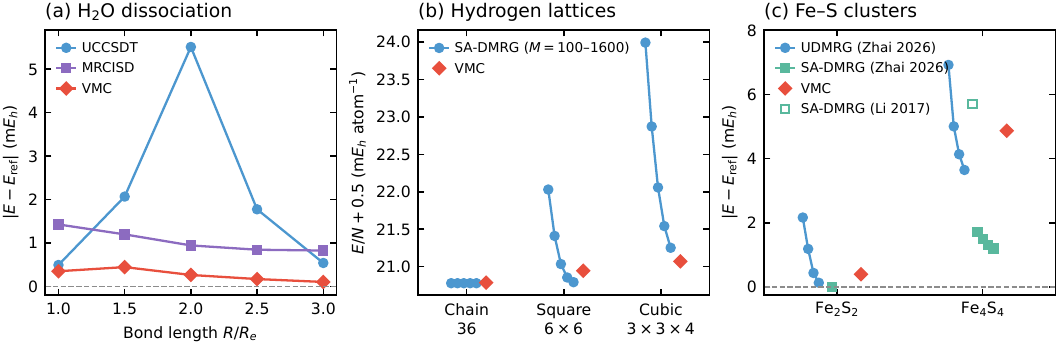}
  \caption{Variational energy benchmarks. (a) Errors relative to FCI for all-electron $(10e,24o)$ \ce{H2O}/cc-pVDZ, with UCCSDT and MRCISD for comparison.\cite{olsen1996full} (b) Energies per atom for half-filled 36-site STO-6G chain, $6\times6$ square, and $3\times3\times4$ cubic lattices, compared with SA-DMRG at $M=100$--$1600$. (c) Errors for \ce{Fe2S2} $(30e,20o)$ and \ce{Fe4S4} $(54e,36o)$. References are $-116.605609\,E_{\mathrm h}$ ($M=12000$ SA-DMRG) and $-327.245342\,E_{\mathrm h}$ (extrapolated BS1 SA-DMRG), respectively. Filled DMRG points are from Ref.~\cite{zhai2026classical}; the open square is from Ref.~\cite{li2017spinprojected}.}
  \label{fig:energy_benchmarks}
\end{figure*}

For \ce{H2O}, the IR-FS-VMC error remains $0.102$--$0.444\,\mathrm{m}E_{\mathrm h}$ from equilibrium to $3R_{\mathrm e}$ [Fig.~\ref{fig:energy_benchmarks}(a)]. At $2R_{\mathrm e}$ it is $0.263\,\mathrm{m}E_{\mathrm h}$, compared with $5.512$ and $0.943\,\mathrm{m}E_{\mathrm h}$ for UCCSDT and MRCISD, respectively.\cite{olsen1996full} Bond stretching changes the occupation structure substantially, but no sampling or optimization parameter is retuned along the curve. The sub-millihartree errors show that the production protocol remains accurate from near equilibrium into the stretched, multireference regime.

The 36-site hydrogen systems extend the comparison to one-, two-, and three-dimensional geometries [Fig.~\ref{fig:energy_benchmarks}(b)]. IR-FS-VMC lies $0.007$ and $0.153\,\mathrm{m}E_{\mathrm h}$ per atom above the $M=1600$ SA-DMRG values for the chain and square lattice and $0.182\,\mathrm{m}E_{\mathrm h}$ per atom below the corresponding cubic value. The SA-DMRG energy is essentially converged over the displayed bond dimensions for the chain, whereas it changes by $1.238$ and $2.741\,\mathrm{m}E_{\mathrm h}$ per atom for the square and cubic lattices between $M=100$ and $1600$. Accordingly, the cubic $M=1600$ point is a finite-bond-dimension variational reference rather than a converged ground-state estimate; an IR-FS-VMC energy below that point is therefore consistent with the variational principle. Across the three geometries, the IR-FS-VMC energies remain within $0.2\,\mathrm{m}E_{\mathrm h}$ per atom of the best displayed SA-DMRG values, while the latter show increasingly strong bond-dimension dependence from the chain to the square and cubic lattices.

The Fe--S models probe a different form of many-electron complexity, with dense low-energy spin couplings and substantial multireference character in compact active spaces.\cite{sharma2014fes,li2017spinprojected,zhai2026classical} For \ce{Fe2S2} $(30e,20o)$, IR-FS-VMC reaches $-116.605210\,E_{\mathrm h}$, $0.399\,\mathrm{m}E_{\mathrm h}$ above the $M=12000$ SA-DMRG reference. For \ce{Fe4S4} $(54e,36o)$, it reaches $-327.240478\,E_{\mathrm h}$, $4.864\,\mathrm{m}E_{\mathrm h}$ above the extrapolated BS1 SA-DMRG reference.\cite{zhai2026classical} Taken together with \ce{H2O} and \ce{H36}, the energy benchmarks show that IR-FS-VMC can be used without system-specific sampling parameters across markedly different electronic structures.

\subsection{Electronic Correlation}
\label{sec:wavefunction_structure}

Natural occupations and spin correlations characterize the distinct correlation regimes represented in the benchmark set.

\begin{figure*}[t]
  \centering
  \includegraphics[width=\textwidth]{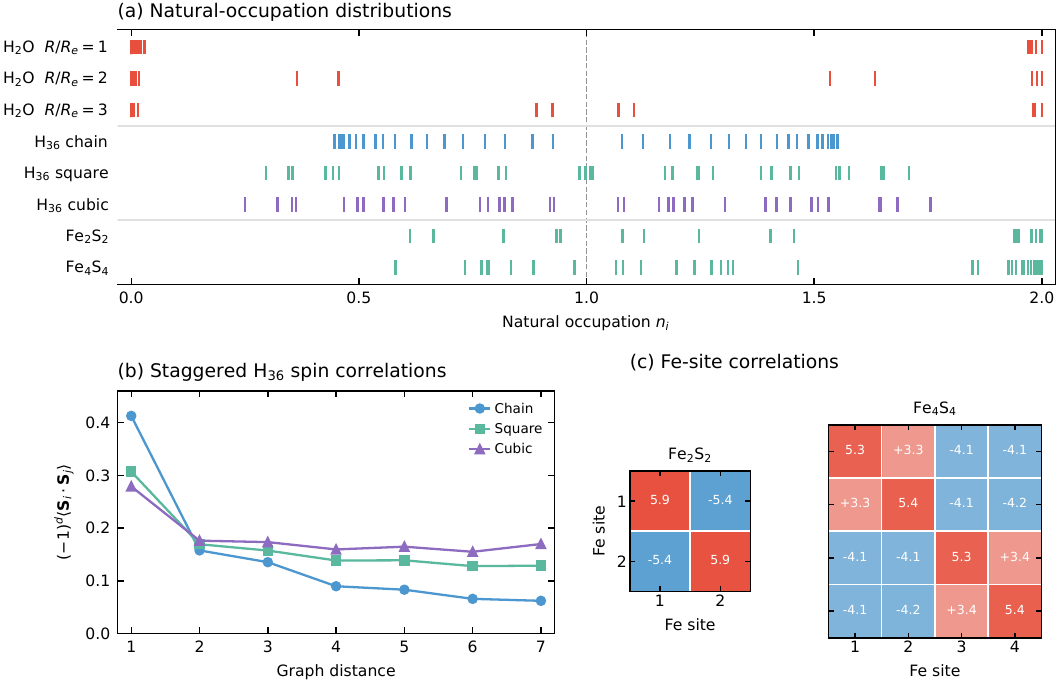}
  \caption{Electronic-correlation diagnostics for the optimized IR-FS-VMC states. (a) Natural occupations for \ce{H2O} at $R/R_{\mathrm e}=1$, 2, and 3, the three \ce{H36} lattices, and the iron--sulfur clusters; the dashed line marks $n_i=1$. (b) Staggered \ce{H36} spin correlations $(-1)^d\langle\bm S_i\!\cdot\!\bm S_j\rangle$, averaged over orbital pairs at Manhattan distance $d$. (c) Fe-site correlations $\langle\bm S_I\!\cdot\!\bm S_J\rangle$, obtained by summing over the five Fe $3d$ orbitals on each site; diagonal entries are $\langle\bm S_I^2\rangle$.}
  \label{fig:wavefunction_structure}
\end{figure*}

The natural occupations span qualitatively different regimes [Fig.~\ref{fig:wavefunction_structure}(a)]. Equilibrium \ce{H2O} is close to a closed-shell $2/0$ occupation pattern, whereas bond stretching produces fractional frontier occupations; at $3R_{\mathrm e}$ four orbitals lie near single occupation. The hydrogen lattices and Fe--S active spaces show broadly fractional spectra, consistent with substantial multireference character.

The spin correlations show the corresponding magnetic structure [Fig.~\ref{fig:wavefunction_structure}(b,c)]. The hydrogen lattices exhibit alternating antiferromagnetic correlations over several Manhattan distances. In \ce{Fe2S2}, the two Fe centers are strongly antiferromagnetically correlated, $\langle\bm S_1\!\cdot\!\bm S_2\rangle=-5.35$. In \ce{Fe4S4}, Fe1--Fe2 and Fe3--Fe4 have positive correlations of about $+3.34$, while correlations between the two pairs are about $-4.15$, consistent with the BS1 spin-coupling pattern used to initialize the calculation and with the strongly spin-coupled character of these active-space models.\cite{sharma2014fes,zhai2026classical}

\section{Conclusions and Outlook}

We have developed IR-FS-VMC, an importance-sampling formulation of FS-VMC in which an evaluation kernel maps the invariant distribution of a Metropolis--Hastings chain to the importance distribution used to estimate the Born-distribution energy, gradient, and SR matrix.

Controlled \ce{H2O} and \ce{Fe2S2} calculations characterize the effects of the sampling and estimation procedure, while the broader benchmarks span near-closed-shell, bond-breaking, extended multireference, and strongly spin-coupled regimes. A single production protocol remains accurate across these systems without system-specific sampling parameters.

The formulation requires only wave-function amplitudes, retained Hamiltonian connections, and the local quantities already used in FS-VMC. It does not require an autoregressive factorization and remains compatible with general amplitude-evaluable ansatzes, optimization methods, and local-energy approximations.

Three directions are especially relevant for improving overall efficiency. First, reducing the formal complexity and practical cost of local-energy evaluation is important for larger systems.\cite{liu2025efficient} Second, adaptive multiple importance sampling suggests reusing configurations across nearby optimization steps through reweighting across multiple distributions.\cite{cornuet2012amis} Third, the importance distribution and proposal kernel can be designed more systematically to improve statistical efficiency at fixed computational cost.\cite{trail2010optimum,misery2026importance}

\section*{SUPPLEMENTARY MATERIAL}
The supplementary material provides estimator properties and support conditions, sampling and optimization details, projected neural backflow and semistochastic local-energy definitions, computational specifications, and post-optimization analysis.

\begin{acknowledgments}
The author acknowledges the Supercomputing Center of the University of Science and Technology of China for providing the computational resources used in this work. Generative AI was used for language polishing and text refinement. The author also thanks Qiang Fu, Jiace Sun, and Lixue Cheng for their support.
\end{acknowledgments}

\section*{AUTHOR DECLARATIONS}

\subsection*{Conflict of Interest}
The author has no conflicts to disclose.

\subsection*{Author Contributions}
Zheng Che: Conceptualization; Formal analysis; Investigation; Methodology; Software; Validation; Visualization; Writing -- original draft; Writing -- review \& editing.

\section*{DATA AVAILABILITY}
The data that support the findings of this study are openly available at \url{https://github.com/wsmxcz/IR-FS-VMC-scripts-2026}. The source code is openly available at \url{https://github.com/wsmxcz/FockVMC}.

\bibliography{reference}

@article{holmes2016heatbath,
  author  = {Holmes, Adam A. and Tubman, Norm M. and Umrigar, C. J.},
  title   = {Heat-bath configuration interaction: An efficient selected configuration interaction algorithm inspired by heat-bath sampling},
  journal = {J. Chem. Theory Comput.},
  volume  = {12},
  number  = {8},
  pages   = {3674--3680},
  year    = {2016},
  doi     = {10.1021/acs.jctc.6b00407}
}

@article{bartlett2007coupled,
  author  = {Bartlett, Rodney J. and Musia{\l}, Monika},
  title   = {Coupled-cluster theory in quantum chemistry},
  journal = {Rev. Mod. Phys.},
  volume  = {79},
  number  = {1},
  pages   = {291--352},
  year    = {2007},
  doi     = {10.1103/RevModPhys.79.291}
}

@article{chan2011density,
  author  = {Chan, Garnet Kin-Lic and Sharma, Sandeep},
  title   = {The density matrix renormalization group in quantum chemistry},
  journal = {Annu. Rev. Phys. Chem.},
  volume  = {62},
  pages   = {465--481},
  year    = {2011},
  doi     = {10.1146/annurev-physchem-032210-103338}
}

@article{booth2009fermion,
  author  = {Booth, George H. and Thom, Alex J. W. and Alavi, Ali},
  title   = {Fermion {Monte Carlo} without fixed nodes: A game of life, death, and annihilation in {Slater} determinant space},
  journal = {J. Chem. Phys.},
  volume  = {131},
  number  = {5},
  pages   = {054106},
  year    = {2009},
  doi     = {10.1063/1.3193710}
}

@article{foulkes2001quantum,
  author  = {Foulkes, W. M. C. and Mitas, Lubos and Needs, R. J. and Rajagopal, G.},
  title   = {Quantum {Monte Carlo} simulations of solids},
  journal = {Rev. Mod. Phys.},
  volume  = {73},
  number  = {1},
  pages   = {33--83},
  year    = {2001},
  doi     = {10.1103/RevModPhys.73.33}
}

@article{choo2020fermionic,
  author  = {Choo, Kenny and Mezzacapo, Antonio and Carleo, Giuseppe},
  title   = {Fermionic neural-network states for ab-initio electronic structure},
  journal = {Nat. Commun.},
  volume  = {11},
  number  = {1},
  pages   = {2368},
  year    = {2020},
  doi     = {10.1038/s41467-020-15724-9}
}

@article{barrett2022autoregressive,
  author  = {Barrett, Thomas D. and Malyshev, Aleksei and Lvovsky, A. I.},
  title   = {Autoregressive neural-network wavefunctions for ab initio quantum chemistry},
  journal = {Nat. Mach. Intell.},
  volume  = {4},
  number  = {4},
  pages   = {351--358},
  year    = {2022},
  doi     = {10.1038/s42256-022-00461-z}
}

@article{zhao2023scalable,
  author  = {Zhao, Tianchen and Stokes, James and Veerapaneni, Shravan},
  title   = {Scalable neural quantum states architecture for quantum chemistry},
  journal = {Mach. Learn.: Sci. Technol.},
  volume  = {4},
  number  = {2},
  pages   = {025034},
  year    = {2023},
  doi     = {10.1088/2632-2153/acdb2f}
}

@article{carleo2017solving,
  author  = {Carleo, Giuseppe and Troyer, Matthias},
  title   = {Solving the quantum many-body problem with artificial neural networks},
  journal = {Science},
  volume  = {355},
  number  = {6325},
  pages   = {602--606},
  year    = {2017},
  doi     = {10.1126/science.aag2302}
}

@article{holmes2016efficient,
  author  = {Holmes, Adam A. and Changlani, Hitesh J. and Umrigar, C. J.},
  title   = {Efficient heat-bath sampling in {Fock space}},
  journal = {J. Chem. Theory Comput.},
  volume  = {12},
  number  = {4},
  pages   = {1561--1571},
  year    = {2016},
  doi     = {10.1021/acs.jctc.5b01170}
}

@article{sorella2001generalized,
  author  = {Sorella, Sandro},
  title   = {Generalized {Lanczos} algorithm for variational quantum {Monte Carlo}},
  journal = {Phys. Rev. B},
  volume  = {64},
  number  = {2},
  pages   = {024512},
  year    = {2001},
  doi     = {10.1103/PhysRevB.64.024512}
}

@article{stokes2020quantum,
  author  = {Stokes, James and Izaac, Josh and Killoran, Nathan and Carleo, Giuseppe},
  title   = {Quantum natural gradient},
  journal = {Quantum},
  volume  = {4},
  pages   = {269},
  year    = {2020},
  doi     = {10.22331/q-2020-05-25-269}
}

@article{trail2008alternative,
  author  = {Trail, J. R.},
  title   = {Alternative sampling for variational quantum {Monte Carlo}},
  journal = {Phys. Rev. E},
  volume  = {77},
  number  = {1},
  pages   = {016704},
  year    = {2008},
  doi     = {10.1103/PhysRevE.77.016704}
}

@article{trail2010optimum,
  author  = {Trail, J. R. and Maezono, Ryo},
  title   = {Optimum and efficient sampling for variational quantum {Monte Carlo}},
  journal = {J. Chem. Phys.},
  volume  = {133},
  number  = {17},
  pages   = {174120},
  year    = {2010},
  doi     = {10.1063/1.3488651}
}

@article{wan2026removing,
  author  = {Wan, Zhou-Quan and Wiersema, Roeland and Zhang, Shiwei},
  title   = {Removing nodal and support-mismatch pathologies in variational {Monte Carlo} via blurred sampling},
  journal = {Phys. Rev. X},
  year    = {2026},
  volume  = {16},
  number  = {3},
  pages   = {031059},
  doi     = {10.1103/jrn5-gv19}
}

@article{luo2019backflow,
  author  = {Luo, Di and Clark, Bryan K.},
  title   = {Backflow transformations via neural networks for quantum many-body wave functions},
  journal = {Phys. Rev. Lett.},
  volume  = {122},
  number  = {22},
  pages   = {226401},
  year    = {2019},
  doi     = {10.1103/PhysRevLett.122.226401}
}

@article{sorella1998stochastic,
  author  = {Sorella, Sandro},
  title   = {Green function {Monte Carlo} with stochastic reconfiguration},
  journal = {Phys. Rev. Lett.},
  volume  = {80},
  number  = {20},
  pages   = {4558--4561},
  year    = {1998},
  doi     = {10.1103/PhysRevLett.80.4558}
}

@article{chen2024minsr,
  author  = {Chen, Ao and Heyl, Markus},
  title   = {Empowering deep neural quantum states through efficient optimization},
  journal = {Nat. Phys.},
  volume  = {20},
  number  = {9},
  pages   = {1476--1481},
  year    = {2024},
  doi     = {10.1038/s41567-024-02566-1}
}

@article{goldshlager2024spring,
  author  = {Goldshlager, Gil and Abrahamsen, Nilin and Lin, Lin},
  title   = {A {Kaczmarz}-inspired approach to accelerate the optimization of neural network wavefunctions},
  journal = {J. Comput. Phys.},
  volume  = {516},
  pages   = {113351},
  year    = {2024},
  doi     = {10.1016/j.jcp.2024.113351}
}

@article{sabzevari2018orbital,
  author  = {Sabzevari, Iliya and Sharma, Sandeep},
  title   = {Improved speed and scaling in orbital space variational {Monte Carlo}},
  journal = {J. Chem. Theory Comput.},
  volume  = {14},
  number  = {12},
  pages   = {6276--6286},
  year    = {2018},
  doi     = {10.1021/acs.jctc.8b00780}
}

@article{li2023nonstochastic,
  author  = {Li, Xiang and Huang, Jia-Cheng and Zhang, Guang-Ze and Li, Hao-En and Cao, Chang-Su and Lv, Dingshun and Hu, Han-Shi},
  title   = {A nonstochastic optimization algorithm for neural-network quantum states},
  journal = {J. Chem. Theory Comput.},
  volume  = {19},
  number  = {22},
  pages   = {8156--8165},
  year    = {2023},
  doi     = {10.1021/acs.jctc.3c00831}
}

@article{liu2024backflow,
  author  = {Liu, An-Jun and Clark, Bryan K.},
  title   = {Neural network backflow for ab initio quantum chemistry},
  journal = {Phys. Rev. B},
  volume  = {110},
  number  = {11},
  pages   = {115137},
  year    = {2024},
  doi     = {10.1103/PhysRevB.110.115137}
}

@article{misery2026importance,
  author  = {Misery, Antoine and Gravina, Luca and Santini, Alessandro and Vicentini, Filippo},
  title   = {Looking elsewhere: Improving variational {Monte Carlo} gradients by importance sampling},
  journal = {Mach. Learn.: Sci. Technol.},
  volume  = {7},
  number  = {1},
  pages   = {015035},
  year    = {2026},
  doi     = {10.1088/2632-2153/ae387f}
}

@article{yokoyama1987variational,
  author  = {Yokoyama, Hisatoshi and Shiba, Hiroyuki},
  title   = {Variational {Monte-Carlo} studies of {Hubbard} model. I},
  journal = {J. Phys. Soc. Jpn.},
  volume  = {56},
  number  = {4},
  pages   = {1490--1506},
  year    = {1987},
  doi     = {10.1143/JPSJ.56.1490}
}

@article{neuscamman2012optimizing,
  author  = {Neuscamman, Eric and Umrigar, C. J. and Chan, Garnet Kin-Lic},
  title   = {Optimizing large parameter sets in variational quantum {Monte Carlo}},
  journal = {Phys. Rev. B},
  volume  = {85},
  number  = {4},
  pages   = {045103},
  year    = {2012},
  doi     = {10.1103/PhysRevB.85.045103}
}

@article{peng2025optimization,
  author  = {Peng, Ruojing and Chan, Garnet Kin-Lic},
  title   = {First- and quasi-second-order optimization algorithms in variational {Monte Carlo}},
  journal = {Phys. Rev. Res.},
  volume  = {7},
  number  = {4},
  pages   = {043351},
  year    = {2025},
  doi     = {10.1103/r7c3-zq4b}
}

@article{sandvik2007variational,
  author  = {Sandvik, Anders W. and Vidal, Guifr{\'e}},
  title   = {Variational quantum {Monte Carlo} simulations with tensor-network states},
  journal = {Phys. Rev. Lett.},
  volume  = {99},
  number  = {22},
  pages   = {220602},
  year    = {2007},
  doi     = {10.1103/PhysRevLett.99.220602}
}

@article{nomura2017restricted,
  author  = {Nomura, Yusuke and Darmawan, Andrew S. and Yamaji, Youhei and Imada, Masatoshi},
  title   = {Restricted {Boltzmann} machine learning for solving strongly correlated quantum systems},
  journal = {Phys. Rev. B},
  volume  = {96},
  number  = {20},
  pages   = {205152},
  year    = {2017},
  doi     = {10.1103/PhysRevB.96.205152}
}

@article{wei2018reduced,
  author  = {Wei, Haochuan and Neuscamman, Eric},
  title   = {Reduced scaling Hilbert space variational quantum {Monte Carlo}},
  journal = {J. Chem. Phys.},
  volume  = {149},
  number  = {18},
  pages   = {184106},
  year    = {2018},
  doi     = {10.1063/1.5047207}
}

@article{che2026deterministic,
  author  = {Che, Zheng},
  title   = {A deterministic framework for neural network quantum states in quantum chemistry},
  journal = {J. Chem. Theory Comput.},
  volume  = {22},
  number  = {13},
  pages   = {6497--6509},
  year    = {2026},
  doi     = {10.1021/acs.jctc.6c00445}
}

@article{tubman2016deterministic,
  author  = {Tubman, Norm M. and Lee, Joonho and Takeshita, Tyler Y. and Head-Gordon, Martin and Whaley, K. Birgitta},
  title   = {A deterministic alternative to the full configuration interaction quantum {Monte Carlo} method},
  journal = {J. Chem. Phys.},
  volume  = {145},
  number  = {4},
  pages   = {044112},
  year    = {2016},
  doi     = {10.1063/1.4955109}
}

@article{sharma2017semistochastic,
  author  = {Sharma, Sandeep and Holmes, Adam A. and Jeanmairet, Guillaume and Alavi, Ali and Umrigar, C. J.},
  title   = {Semistochastic heat-bath configuration interaction method: Selected configuration interaction with semistochastic perturbation theory},
  journal = {J. Chem. Theory Comput.},
  volume  = {13},
  number  = {4},
  pages   = {1595--1604},
  year    = {2017},
  doi     = {10.1021/acs.jctc.6b01028}
}

@article{chan2002dmrg,
  author  = {Chan, Garnet Kin-Lic and Head-Gordon, Martin},
  title   = {Highly correlated calculations with a polynomial cost algorithm: A study of the density matrix renormalization group},
  journal = {J. Chem. Phys.},
  volume  = {116},
  number  = {11},
  pages   = {4462--4476},
  year    = {2002},
  doi     = {10.1063/1.1449459}
}

@article{sharma2012spinadapted,
  author  = {Sharma, Sandeep and Chan, Garnet Kin-Lic},
  title   = {Spin-adapted density matrix renormalization group algorithms for quantum chemistry},
  journal = {J. Chem. Phys.},
  volume  = {136},
  number  = {12},
  pages   = {124121},
  year    = {2012},
  doi     = {10.1063/1.3695642}
}

@article{tahara2008variational,
  author  = {Tahara, Daisuke and Imada, Masatoshi},
  title   = {Variational {Monte Carlo} method combined with quantum-number projection and multi-variable optimization},
  journal = {J. Phys. Soc. Jpn.},
  volume  = {77},
  number  = {11},
  pages   = {114701},
  year    = {2008},
  doi     = {10.1143/JPSJ.77.114701}
}

@article{choo2019j1j2,
  author  = {Choo, Kenny and Neupert, Titus and Carleo, Giuseppe},
  title   = {Two-dimensional frustrated {$J_1$--$J_2$} model studied with neural network quantum states},
  journal = {Phys. Rev. B},
  volume  = {100},
  number  = {12},
  pages   = {125124},
  year    = {2019},
  doi     = {10.1103/PhysRevB.100.125124}
}

@article{viteritti2023transformer,
  author  = {Viteritti, Luciano Loris and Rende, Riccardo and Becca, Federico},
  title   = {Transformer variational wave functions for frustrated quantum spin systems},
  journal = {Phys. Rev. Lett.},
  volume  = {130},
  number  = {23},
  pages   = {236401},
  year    = {2023},
  doi     = {10.1103/PhysRevLett.130.236401}
}

@article{olsen1996full,
  author  = {Olsen, Jeppe and J{\o}rgensen, Poul and Koch, Henrik and Balkova, Anna and Bartlett, Rodney J.},
  title   = {Full configuration-interaction and state of the art correlation calculations on water in a valence double-zeta basis with polarization functions},
  journal = {J. Chem. Phys.},
  volume  = {104},
  number  = {20},
  pages   = {8007--8015},
  year    = {1996},
  doi     = {10.1063/1.471518}
}

@article{li2017spinprojected,
  author  = {Li, Zhendong and Chan, Garnet Kin-Lic},
  title   = {Spin-projected matrix product states: Versatile tool for strongly correlated systems},
  journal = {J. Chem. Theory Comput.},
  volume  = {13},
  number  = {6},
  pages   = {2681--2695},
  year    = {2017},
  doi     = {10.1021/acs.jctc.7b00270}
}

@misc{zhai2026classical,
  author  = {Zhai, Huanchen and Li, Chenghan and Zhang, Xing and Li, Zhendong and Lee, Seunghoon and Chan, Garnet Kin-Lic},
  title   = {Classical computational simulation of the {FeMo}-cofactor model to chemical accuracy and its implications},
  year    = {2026},
  note    = {arXiv:2601.04621v2},
  doi     = {10.48550/arXiv.2601.04621}
}

@article{petruzielo2012semistochastic,
  author  = {Petruzielo, F. R. and Holmes, A. A. and Changlani, Hitesh J. and Nightingale, M. P. and Umrigar, C. J.},
  title   = {Semistochastic projector {Monte Carlo} method},
  journal = {Phys. Rev. Lett.},
  volume  = {109},
  number  = {23},
  pages   = {230201},
  year    = {2012},
  doi     = {10.1103/PhysRevLett.109.230201}
}

@article{wu2025hybrid,
  author  = {Wu, Zibo and Zhang, Bohan and Fang, Wei-Hai and Li, Zhendong},
  title   = {Hybrid tensor network and neural network quantum states for quantum chemistry},
  journal = {J. Chem. Theory Comput.},
  volume  = {21},
  number  = {20},
  pages   = {10252--10262},
  year    = {2025},
  doi     = {10.1021/acs.jctc.5c01228}
}

@article{rath2023framework,
  author  = {Rath, Yannic and Booth, George H.},
  title   = {Framework for efficient ab initio electronic structure with {Gaussian Process States}},
  journal = {Phys. Rev. B},
  volume  = {107},
  number  = {20},
  pages   = {205119},
  year    = {2023},
  doi     = {10.1103/PhysRevB.107.205119}
}

@article{geweke1989bayesian,
  author  = {Geweke, John},
  title   = {Bayesian inference in econometric models using {Monte Carlo} integration},
  journal = {Econometrica},
  volume  = {57},
  number  = {6},
  pages   = {1317--1339},
  year    = {1989},
  doi     = {10.2307/1913710}
}

@article{tierney1994markov,
  author  = {Tierney, Luke},
  title   = {Markov chains for exploring posterior distributions},
  journal = {Ann. Stat.},
  volume  = {22},
  number  = {4},
  pages   = {1701--1728},
  year    = {1994},
  doi     = {10.1214/aos/1176325750}
}

@article{zhai2023block2,
  author  = {Zhai, Huanchen and Larsson, Henrik R. and Lee, Seunghoon and Cui, Zhi-Hao and Zhu, Tianyu and Sun, Chong and Peng, Linqing and Peng, Ruojing and Liao, Ke and T{\"o}lle, Johannes and Yang, Junjie and Li, Shuoxue and Chan, Garnet Kin-Lic},
  title   = {{Block2}: A comprehensive open source framework to develop and apply state-of-the-art {DMRG} algorithms in electronic structure and beyond},
  journal = {J. Chem. Phys.},
  volume  = {159},
  number  = {23},
  pages   = {234801},
  year    = {2023},
  doi     = {10.1063/5.0180424}
}

@article{geyer1992practical,
  author  = {Geyer, Charles J.},
  title   = {Practical Markov chain {Monte Carlo}},
  journal = {Stat. Sci.},
  volume  = {7},
  number  = {4},
  pages   = {473--483},
  year    = {1992},
  doi     = {10.1214/ss/1177011137}
}

@article{sun2020pyscf,
  author  = {Sun, Qiming and
             Zhang, Xing and
             Banerjee, Samragni and
             Bao, Peng and
             Barbry, Marc and
             Blunt, Nick S. and
             Bogdanov, Nikolay A. and
             Booth, George H. and
             Chen, Jia and
             Cui, Zhi-Hao and
             Eriksen, Janus J. and
             Gao, Yang and
             Guo, Sheng and
             Hermann, Jan and
             Hermes, Matthew R. and
             Koh, Kevin and
             Koval, Peter and
             Lehtola, Susi and
             Li, Zhendong and
             Liu, Junzi and
             Mardirossian, Narbe and
             McClain, James D. and
             Motta, Mario and
             Mussard, Bastien and
             Pham, Hung Q. and
             Pulkin, Artem and
             Purwanto, Wirawan and
             Robinson, Paul J. and
             Ronca, Enrico and
             Sayfutyarova, Elvira R. and
             Scheurer, Maximilian and
             Schurkus, Henry F. and
             Smith, James E. T. and
             Sun, Chong and
             Sun, Shi-Ning and
             Upadhyay, Shiv and
             Wagner, Lucas K. and
             Wang, Xiao and
             White, Alec and
             Whitfield, James Daniel and
             Williamson, Mark J. and
             Wouters, Sebastian and
             Yang, Jun and
             Yu, Jason M. and
             Zhu, Tianyu and
             Berkelbach, Timothy C. and
             Sharma, Sandeep and
             Sokolov, Alexander Yu. and
             Chan, Garnet Kin-Lic},
  title   = {Recent developments in the {PySCF} program package},
  journal = {J. Chem. Phys.},
  volume  = {153},
  number  = {2},
  pages   = {024109},
  year    = {2020},
  doi     = {10.1063/5.0006074}
}

@article{liu2025efficient,
  author  = {Liu, An-Jun and Clark, Bryan K.},
  title   = {Efficient optimization of neural network backflow for ab initio quantum chemistry},
  journal = {Phys. Rev. B},
  volume  = {112},
  number  = {15},
  pages   = {155162},
  year    = {2025},
  doi     = {10.1103/thz7-lmdn}
}

@misc{li2026spinadapted,
  author  = {Li, Yunzhi and Wu, Zibo and Zhang, Bohan and Fang, Wei-Hai and Li, Zhendong},
  title   = {Spin-adapted neural network backflow for strongly correlated electrons},
  year    = {2026},
  note    = {arXiv:2604.06841},
  doi     = {10.48550/arXiv.2604.06841}
}

@article{solanki2026selected,
  author  = {Solanki, Marco Julian and Ding, Lexin and Reiher, Markus},
  title   = {Neural quantum states based on selected configurations},
  journal = {J. Phys. Chem. Lett.},
  volume  = {17},
  number  = {18},
  pages   = {5180--5190},
  year    = {2026},
  doi     = {10.1021/acs.jpclett.6c00520}
}

@article{sharma2014fes,
  author  = {Sharma, Sandeep and Sivalingam, Kantharuban and Neese, Frank and Chan, Garnet Kin-Lic},
  title   = {Low-energy spectrum of iron-sulfur clusters directly from many-particle quantum mechanics},
  journal = {Nat. Chem.},
  volume  = {6},
  number  = {10},
  pages   = {927--933},
  year    = {2014},
  doi     = {10.1038/nchem.2041}
}

@article{sun2018pyscf,
  author  = {Sun, Qiming and
             Berkelbach, Timothy C. and
             Blunt, Nick S. and
             Booth, George H. and
             Guo, Sheng and
             Li, Zhendong and
             Liu, Junzi and
             McClain, James D. and
             Sayfutyarova, Elvira R. and
             Sharma, Sandeep and
             Wouters, Sebastian and
             Chan, Garnet Kin-Lic},
  title   = {{PySCF}: The Python-based simulations of chemistry framework},
  journal = {Wiley Interdiscip. Rev.: Comput. Mol. Sci.},
  volume  = {8},
  number  = {1},
  pages   = {e1340},
  year    = {2018},
  doi     = {10.1002/wcms.1340}
}

@article{cornuet2012amis,
  author  = {Cornuet, Jean-Marie and Marin, Jean-Michel and Mira, Antonietta and Robert, Christian P.},
  title   = {Adaptive Multiple Importance Sampling},
  journal = {Scand. J. Stat.},
  volume  = {39},
  number  = {4},
  pages   = {798--812},
  year    = {2012},
  doi     = {10.1111/j.1467-9469.2011.00756.x}
}

\end{document}


\title{Supplementary Material: Importance-Reweighted Fock-Space Variational Monte Carlo}

\author{Zheng Che}
\email{wsmxcz@gmail.com}
\affiliation{Hefei National Research Center for Physical Sciences at the Microscale, University of Science and Technology of China, Hefei 230026, China}

\maketitle

\section{Importance-Reweighted Estimators}
\label{sec:estimator_details}

We use the notation of the main text and suppress parameter dependence where unambiguous. This section gives the statistical properties of the self-normalized importance-sampling estimators.

\subsection{Self-Normalized Estimation}
\label{sec:si_reweighted_expectations}

For samples from the importance distribution $r$, the normalized importance weights $w_n$ are those defined in the main text. Repeated configurations are combined before wave-function evaluation. If a configuration $x$ occurs $N_x$ times, its contribution to the unnormalized weight is
\begin{equation}
N_x\exp\!\left[2\log|\psi(x)|-\log\widetilde r(x)\right],
\label{eq:si_aggregated_weight}
\end{equation}
which is algebraically identical to assigning the same importance weight to every occurrence.

The same normalized importance weights apply to any operator $\hat A$ whose local estimator is $A_{\mathrm L}(x)=\langle x|\hat A|\psi\rangle/\psi(x)$, giving $\widehat A=\sum_n w_n A_{\mathrm L}(x_n)$. As usual for self-normalized importance sampling, such self-normalized estimators are generally biased at finite sample size but are consistent under the corresponding ergodicity and moment conditions.\cite{geweke1989bayesian,trail2008alternative,trail2010optimum}

\subsection{Exact Marginalization}
\label{sec:si_marginal_construction}

Let the Markov chain have invariant distribution $\rho_\alpha$, and let $K_\beta$ be the evaluation kernel defined in the main text. The joint distribution of the chain configuration $x$ and evaluation configuration $y$ is $\rho_\alpha(x)K_\beta(y|x)$, and marginalizing over the chain configuration gives
\begin{equation}
r_{\alpha,\beta}(y)=\sum_x\rho_\alpha(x)K_\beta(y|x).
\label{eq:si_marginal_definition}
\end{equation}
Because $K_\beta$ is normalized, the unnormalized importance distribution has the same normalization constant as the unnormalized invariant distribution,
\begin{equation}
Z_r=\sum_y\widetilde r_{\alpha,\beta}(y)
=\sum_x d(x)|\psi(x)|^\alpha
=Z_\rho.
\label{eq:si_marginal_normalization}
\end{equation}
Thus $Z_r=Z_\rho$ independently of $\beta$, and the importance weight is $|\psi(y)|^2/\widetilde r_{\alpha,\beta}(y)$.

\subsection{Asymptotic Variance}
\label{sec:si_statistical_efficiency}

Let $\mu_f=\mathbb E_p[f]$ and $h_f(y)=\omega(y)[f(y)-\mu_f]$. For independent samples from $r$, the self-normalized estimator has asymptotic variance\cite{geweke1989bayesian}
\begin{equation}
\sigma_{f,\mathrm{iid}}^2
=\frac{\mathbb E_r[\omega^2(f-\mu_f)^2]}
{\mathbb E_r[\omega]^2},
\label{eq:si_snis_variance}
\end{equation}
whereas for a stationary correlated evaluation sequence $\{Y_n\}$,\cite{tierney1994markov}
\begin{align}
\sigma_f^2
&=\frac{\gamma_f(0)+2\sum_{k\ge1}\gamma_f(k)}{\mathbb E_r[\omega]^2},
\label{eq:si_snis_longrun}\\
\gamma_f(k)&=\operatorname{Cov}[h_f(Y_0),h_f(Y_k)].
\end{align}
The $k=0$ term is the variance of $h_f$, whereas the $k>0$ terms are autocovariances inherited from the Markov chain. Thus the estimator variance depends on both importance weighting and serial correlation.

For the Born-distribution MCMC diagnostics below, we estimate $\tau_{\mathrm{int}}$ for $2\log|\psi|$ with Geyer's initial monotone sequence estimator\cite{geyer1992practical} and report
\begin{equation}
\mathrm{ESS}_{\mathrm{MCMC}}/N=1/\tau_{\mathrm{int}}.
\label{eq:si_mcmc_ess}
\end{equation}
This effective sample size characterizes serial correlation in the Markov chain.

\subsection{Support and Wave-Function Zeros}
\label{sec:si_zeros}

Importance sampling requires $r(x)>0$ wherever $p(x)>0$. For $0\le\beta<1$ and $d(x)>0$, the direct term in $\widetilde r_{\alpha,\beta}$ gives
\begin{equation}
\widetilde r_{\alpha,\beta}(x)
\ge (1-\beta)d(x)|\psi(x)|^\alpha,
\qquad
0\le\omega(x)
\le\frac{|\psi(x)|^{2-\alpha}}
{(1-\beta)d(x)},
\quad 0\le\alpha\le2.
\label{eq:si_weight_bound}
\end{equation}
Thus the support of the importance distribution contains the support of $p$ whenever $d(x)>0$. If $d(x)=0$, no proposed configuration is generated; the current chain configuration is retained, with $|\psi(x)|^\alpha$ as the stay contribution to the unnormalized importance distribution.

Quantities involving division by the amplitude, including $E_{\mathrm L}(x)=(H\psi)(x)/\psi(x)$ and $O_i(x)=\partial_{\theta_i}\psi(x)/\psi(x)$, are undefined at $\psi(x)=0$. In discrete configuration spaces this is closely related to the support-mismatch problem discussed in recent VMC work.\cite{wan2026removing} We therefore assume nonzero amplitudes on configurations where these quantities are evaluated.

\section{Sampling, Estimation, and Optimization}
\label{sec:si_sampling_optimization}

\subsection{Metropolis--Hastings Sampling}
\label{sec:si_markov_chain}

The Markov chain has invariant distribution $\rho_\alpha(x)\propto d(x)|\psi(x)|^\alpha$ and uses the Metropolis--Hastings proposal kernel $Q_H(y|x)=|H_{xy}|/d(x)$. The proposal probabilities are proportional to $|H_{xy}|$ over the retained connections.\cite{booth2009fermion,holmes2016efficient} Production uses the deterministic strong-connection set of the semistochastic local-energy evaluation as $\mathcal C(x)$, consisting of retained single and double excitations constructed with integral-based screening.

For a retained pair with nonzero amplitudes, Hermiticity reduces the Metropolis--Hastings ratio to
\begin{equation}
\frac{\rho_\alpha(y)Q_H(x|y)}
{\rho_\alpha(x)Q_H(y|x)}
=
\frac{d(y)|\psi(y)|^\alpha |H_{yx}|/d(y)}
{d(x)|\psi(x)|^\alpha |H_{xy}|/d(x)}
=\left|\frac{\psi(y)}{\psi(x)}\right|^\alpha.
\label{eq:si_acceptance_ratio}
\end{equation}
The acceptance probability is therefore $a(x,y)=\min\{1,|\psi(y)/\psi(x)|^\alpha\}$. The factor $d(x)$ cancels because it appears in both the invariant distribution and the proposal ratio.

\subsection{Evaluation Kernel and Importance Distribution}
\label{sec:si_sampling_flow}

At each VMC iteration, one proposed configuration $y\sim Q_H(\cdot|x)$ is drawn for each observed chain configuration $x$ and used in the Metropolis--Hastings accept/reject step. The proposed configuration is used for evaluation with probability $\beta$, independently of whether it is accepted into the chain; otherwise $x$ is used. This implements $K_\beta$ conditioned on the pre-transition configuration, and marginalization over the invariant distribution gives Eq.~(\ref{eq:si_marginal_definition}). At fixed $\alpha$, $\beta$ does not alter the Metropolis--Hastings transition kernel.

Production calculations use $N_{\mathrm c}=N_{\mathrm s}=4096$. Each optimization iteration records one evaluation configuration from every chain. The proposed configuration associated with that observation also serves as the first of $N_{\mathrm{discard}}=16$ chain transitions; 15 additional transitions are then taken before the next observation. Chains are thermalized once at initialization and then continued across optimization iterations.

For Fig.~2 of the main text, only the sampling and estimation procedure is varied. Born uses the Born distribution with a proposal kernel given by an equal mixture of single- and double-excitation moves and requires no reweighting. Tempered uses an invariant distribution proportional to $|\psi|^\alpha$ with the same proposal kernel and weights proportional to $|\psi|^{2-\alpha}$. The three IR protocols use $(\alpha,\beta)=(2,0.5)$, adaptive $\alpha$ with $\beta=0$, and adaptive $\alpha$ with $\beta=0.5$. The last is the production protocol. The wave function, local-energy evaluation, optimizer, sample size, and initialization are otherwise identical.

\subsection{Adaptive $\alpha$}
\label{sec:si_alpha}

The exponent $\alpha$ controls the amplitude dependence of the invariant distribution. For amplitude-only tempering, its reweighting cost can be written explicitly. Let $Z_s=\sum_x|\psi(x)|^s$ and let $\mathbb E_\alpha[\cdot]$ denote expectation under $|\psi|^\alpha/Z_\alpha$. Then
\begin{equation}
\frac{\mathbb E_\alpha[\omega^2]}{\mathbb E_\alpha[\omega]^2}
=\frac{Z_\alpha Z_{4-\alpha}}{Z_2^2}.
\label{eq:si_alpha_weight_moment}
\end{equation}
If $\log Z_s=Lf(s)+o(L)$ with a strictly convex limiting function $f$ and system-size parameter $L$, then
\begin{equation}
\frac{Z_\alpha Z_{4-\alpha}}{Z_2^2}
=\exp\!\left\{L\left[f(\alpha)+f(4-\alpha)-2f(2)\right]+o(L)\right\},
\label{eq:si_alpha_asymptotic}
\end{equation}
which grows exponentially with system size for any fixed $\alpha\ne2$. Thus fixed-$\alpha$ amplitude tempering can produce increasingly uneven importance weights as the system grows. Equations~\ref{eq:si_alpha_weight_moment} and \ref{eq:si_alpha_asymptotic} apply only to amplitude tempering; the IR importance distribution has a different weight structure.

The exponent is held fixed during each sampling-and-estimation step and updated between optimization iterations. For the current unnormalized importance distribution $\widetilde r_\alpha$, define $s_\alpha(x)=\partial_\alpha\log\widetilde r_\alpha(x)$ and $u_\alpha(x)=\partial_\alpha^2\widetilde r_\alpha(x)/\widetilde r_\alpha(x)$. For unique sampled configurations with multiplicities $N_x$, let $\langle a\rangle_{\mathrm{samp}}=N_{\mathrm s}^{-1}\sum_xN_xa(x)$ and
\begin{equation}
\tau_x=
\frac{w_x|\widehat E_{\mathrm L}(x)-\widehat E|}
{\sum_yw_y|\widehat E_{\mathrm L}(y)-\widehat E|},
\qquad
\kappa_\alpha
=\langle u_\alpha\rangle_{\mathrm{samp}}
-\langle s_\alpha\rangle_{\mathrm{samp}}^2.
\label{eq:si_alpha_weights}
\end{equation}
When $\kappa_\alpha$ is finite and positive, the target value and damped update are
\begin{align}
\alpha^\star
&=\Pi_{[0,2]}\!\left[
\alpha+
\frac{\sum_x\tau_xs_\alpha(x)-\langle s_\alpha\rangle_{\mathrm{samp}}}
{\kappa_\alpha}
\right],
\label{eq:si_alpha_target}\\
\alpha
&\leftarrow\alpha+0.02(\alpha^\star-\alpha).
\label{eq:si_alpha_update}
\end{align}
If the residual normalization or $\kappa_\alpha$ is numerically invalid, $\alpha$ is left unchanged. For amplitude tempering, $\widetilde r_\alpha=|\psi|^\alpha$, so $s_\alpha=\log|\psi|$ and $u_\alpha=(\log|\psi|)^2$; for IR the derivatives are evaluated for the importance distribution defined in the main text. The updated $\alpha$ is used at the next iteration.

\subsection{Stochastic Reconfiguration}
\label{sec:si_sr}

All production calculations use the same importance-weighted stochastic reconfiguration (SR) update.\cite{sorella1998stochastic,sorella2001generalized} Let $J_{ni}=O_i(x_n)$ and $\overline J_i=\sum_nw_nJ_{ni}$, using the logarithmic derivatives defined in the main text. Define
\begin{equation}
\mathcal O_{ni}=\sqrt{w_n}(J_{ni}-\overline J_i),
\qquad
b_n=2\sqrt{w_n}[\widehat E_{\mathrm L}(x_n)-\widehat E].
\label{eq:si_sample_jacobian}
\end{equation}
Then $\widehat S=\mathcal O^{\mathsf T}\mathcal O$ and $\widehat{\bm g}=\mathcal O^{\mathsf T}\bm b$, so shifted SR solves $(\widehat S+\lambda I)\bm d=\widehat{\bm g}$. When the sample dimension is smaller than the parameter dimension,
\begin{equation}
(\mathcal O^{\mathsf T}\mathcal O+\lambda I)^{-1}\mathcal O^{\mathsf T}
=
\mathcal O^{\mathsf T}(\mathcal O\mathcal O^{\mathsf T}+\lambda I)^{-1},
\label{eq:si_sample_space_identity}
\end{equation}
which reduces the linear solve to sample space.\cite{chen2024minsr}

We use a predictive variant motivated by the projected-increment construction of SPRING.\cite{goldshlager2024spring} With the preceding unscaled SR direction $\bm d_{t-1}$, set $\bm p_t=\mu\bm d_{t-1}$ and solve
\begin{align}
(\mathcal O_t\mathcal O_t^{\mathsf T}+\lambda I)\bm a_t
&=\bm b_t-\mathcal O_t\bm p_t,
\label{eq:si_psr_solve}\\
\bm d_t
&=\bm p_t+\mathcal O_t^{\mathsf T}\bm a_t,
\qquad
\Delta\bm\theta_t=-\eta\bm d_t.
\label{eq:si_psr_direction}
\end{align}
Setting $\mu=0$ recovers shifted sample-space SR. The predictor acts on the unscaled SR direction; the learning rate is applied only after the solve. All calculations use double precision.

\section{Wave Function and Local-Energy Evaluation}
\label{sec:si_wavefunction_local_energy}

All calculations use the same real spin-projected neural backflow ansatz, which supplies the amplitudes and logarithmic derivatives used by the estimators above.

\subsection{Effective Particle--Hole Representation}
\label{sec:si_particle_hole}

Let $N_{\mathrm{orb}}$ be the number of spatial orbitals and $N_{\uparrow}\ge N_{\downarrow}$ the electron numbers. We use particles when $N_{\uparrow}+N_{\downarrow}\le N_{\mathrm{orb}}$ and holes otherwise. For occupations $n_{p\uparrow},n_{p\downarrow}\in\{0,1\}$, the effective occupations are
\begin{equation}
(m_{p\uparrow},m_{p\downarrow})=
\begin{cases}
(n_{p\uparrow},n_{p\downarrow}), & N_{\uparrow}+N_{\downarrow}\le N_{\mathrm{orb}},\\[2pt]
(1-n_{p\downarrow},1-n_{p\uparrow}), & N_{\uparrow}+N_{\downarrow}>N_{\mathrm{orb}}.
\end{cases}
\label{eq:si_effective_occupations}
\end{equation}
Accordingly, $(N_{\uparrow}^{\mathrm{eff}},N_{\downarrow}^{\mathrm{eff}})=(N_{\uparrow},N_{\downarrow})$ in the particle representation and $(N_{\mathrm{orb}}-N_{\downarrow},N_{\mathrm{orb}}-N_{\uparrow})$ in the hole representation. The spin projection is preserved, $(N_{\uparrow}^{\mathrm{eff}}-N_{\downarrow}^{\mathrm{eff}})/2=(N_{\uparrow}-N_{\downarrow})/2$, while above half filling the matrix dimension is reduced from $N_{\uparrow}+N_{\downarrow}$ to $N_{\mathrm{eff}}=N_{\uparrow}^{\mathrm{eff}}+N_{\downarrow}^{\mathrm{eff}}$.

\subsection{Neural Backflow Orbitals}
\label{sec:si_backflow_network}

At each spatial orbital we form the spin-scalar inputs
\begin{equation}
 q_p=m_{p\uparrow}+m_{p\downarrow}-1,
 \qquad
 \nu_p=|m_{p\uparrow}-m_{p\downarrow}|.
\label{eq:si_backflow_inputs}
\end{equation}
The concatenated vector $(\bm q,\bm\nu)$ is passed through two fully connected hidden layers of width 256 with SiLU activation, followed by a linear output reshaped to $Y_{\bm\theta}(x)\in\mathbb R^{N_{\mathrm{eff}}\times N_{\mathrm{orb}}}$. This defines configuration-dependent spatial orbitals in the neural-backflow form.\cite{luo2019backflow}

The output bias is initialized from a chosen reference orbital matrix. In the particle representation it contains the occupied reference spin-up and spin-down orbitals. In the hole representation, complete QR decompositions of the particle reference subspaces provide orthonormal complements: the complement of the spin-down reference subspace forms the effective spin-up block and the complement of the spin-up reference subspace forms the effective spin-down block. The output-layer weights are initialized from a normal distribution with standard deviation $10^{-3}$, so the initial network is a small configuration-dependent deformation of this reference.

For configuration $x$, let $\mathcal I_{\uparrow}(x)$ and $\mathcal I_{\downarrow}(x)$ denote the occupied effective spatial-orbital indices. Selecting these columns of $Y_{\bm\theta}(x)$ and ordering rows by effective spin gives
\begin{equation}
Y_{\bm\theta}^{\mathcal I_{\uparrow},\mathcal I_{\downarrow}}(x)
=
\begin{pmatrix}
A_{\bm\theta} & B_{\bm\theta}\\
C_{\bm\theta} & D_{\bm\theta}
\end{pmatrix},
\label{eq:si_backflow_blocks}
\end{equation}
where $A_{\bm\theta}$ and $D_{\bm\theta}$ have dimensions $N_{\uparrow}^{\mathrm{eff}}\times N_{\uparrow}^{\mathrm{eff}}$ and $N_{\downarrow}^{\mathrm{eff}}\times N_{\downarrow}^{\mathrm{eff}}$, respectively.

\subsection{Highest-Weight Spin Projection}
\label{sec:si_spin_projection}

We work in the highest-weight sector $S=M_S=(N_{\uparrow}^{\mathrm{eff}}-N_{\downarrow}^{\mathrm{eff}})/2$. The corresponding one-dimensional spin projector is\cite{tahara2008variational,li2017spinprojected}
\begin{equation}
\hat P^S_{SS}
=\frac{2S+1}{2}
\int_0^\pi\sin\vartheta\,d\vartheta\,
\cos^{2S}\!\left(\frac{\vartheta}{2}\right)
 e^{-i\vartheta\hat S_y}.
\label{eq:si_spin_projector}
\end{equation}
The backflow orbitals are computed from spin-symmetric occupation features and held fixed while the spin projector acts on the spin degrees of freedom. For Gauss--Legendre nodes and weights $(\xi_k,w_k^{\mathrm{GL}})$ on $[-1,1]$, define $t_k=(1+\xi_k)/2$ and $z_k=(1-\xi_k)/(1+\xi_k)$. The projected amplitude is
\begin{align}
\psi_{\bm\theta}(x)
={}&\chi_{\mathrm{ph}}(x)\frac{2S+1}{2}
\sum_{k=1}^{N_q} w_k^{\mathrm{GL}} t_k^{N_{\uparrow}^{\mathrm{eff}}}
\nonumber\\[-2pt]
&\times
\det\!\begin{pmatrix}
A_{\bm\theta} & \sqrt{z_k}\,B_{\bm\theta}\\
-\sqrt{z_k}\,C_{\bm\theta} & D_{\bm\theta}
\end{pmatrix},
\label{eq:si_projected_backflow}
\end{align}
with $N_q=\lfloor(N_{\uparrow}^{\mathrm{eff}}+2)/2\rfloor$. After the above change of variables the integrand is a polynomial of degree at most $N_{\uparrow}^{\mathrm{eff}}$, so this Gauss--Legendre rule evaluates the finite spin projection exactly in exact arithmetic.

For numerical evaluation, when $A$ is nonsingular we use the block-determinant identity
\begin{equation}
\det\!\begin{pmatrix}
A & \sqrt z B\\
-\sqrt z C & D
\end{pmatrix}
=
\det(A)\det(D+zCA^{-1}B).
\label{eq:si_schur}
\end{equation}
An LU factorization of $A$ is reused across quadrature nodes to evaluate $CA^{-1}B$; the signed logarithm of $\det(A)$ is then combined with signed log-sum-exp accumulation of $\det(D+z_kCA^{-1}B)$. This is algebraically equivalent to Eq.~(\ref{eq:si_projected_backflow}) but avoids evaluating the full $N_{\mathrm{eff}}\times N_{\mathrm{eff}}$ determinant at every quadrature point.

In the particle representation $\chi_{\mathrm{ph}}(x)=1$. In the hole representation, let $\mathcal H(x)$ contain the unoccupied spin-orbital indices in zero-based ordering $(0\uparrow,\ldots,(N_{\mathrm{orb}}-1)\uparrow,0\downarrow,\ldots,(N_{\mathrm{orb}}-1)\downarrow)$. The particle--hole phase is
\begin{equation}
\chi_{\mathrm{ph}}(x)
=(-1)^{
\displaystyle
\sum_{h\in\mathcal H(x)}h
-\frac{N_{\mathrm{eff}}(N_{\mathrm{eff}}-1)}{2}
+N_{\uparrow}^{\mathrm{eff}}N_{\downarrow}^{\mathrm{eff}}
}.
\label{eq:si_particle_hole_phase}
\end{equation}

\subsection{Semistochastic Local Energy}
\label{sec:si_semistochastic}

For an evaluation configuration $x$, the exact local energy is
\begin{equation}
E_{\mathrm L}(x)=H_{xx}+\sum_{y\ne x}H_{xy}\frac{\psi(y)}{\psi(x)}.
\label{eq:si_exact_local_energy}
\end{equation}
For a two-body electronic Hamiltonian, the number of single- and double-excitation connections is $O(N_{\mathrm{orb}}^4)$ in the worst case, so direct evaluation can require wave-function amplitudes on the full connected space. During optimization we instead partition the off-diagonal action as
\begin{equation}
\begin{aligned}
\mathcal C_{\mathrm s}(x)&=\{y\ne x:|H_{xy}|\ge\varepsilon_1\},\\
\mathcal C_{\mathrm w}(x)&=\{y\ne x:\varepsilon_2\le|H_{xy}|<\varepsilon_1\},\\
\mathcal C_{\mathrm o}(x)&=\{y\ne x:0<|H_{xy}|<\varepsilon_2\}.
\end{aligned}
\label{eq:si_eloc_sets}
\end{equation}
The deterministic strong set is also used as the support of the Metropolis--Hastings proposal kernel in the production calculations, $\mathcal C(x)=\mathcal C_{\mathrm s}(x)$. Thus $\varepsilon_1$ defines both the strong--weak boundary of the local-energy contraction and the support of the proposal kernel. The lower threshold $\varepsilon_2$ and the weak-sample count $N_{\mathrm{eloc}}$ affect only the local-energy estimator.

Let $E_{\mathrm L,s}$, $E_{\mathrm L,w}$, and $E_{\mathrm L,o}$ denote the corresponding off-diagonal contributions, so that
\begin{equation}
E_{\mathrm L}(x)=H_{xx}+E_{\mathrm L,s}(x)
+E_{\mathrm L,w}(x)+E_{\mathrm L,o}(x).
\label{eq:si_eloc_partition}
\end{equation}
The strong contribution is evaluated deterministically. The weak contribution is estimated by stratified importance sampling, following the usual semistochastic separation of dominant and residual Hamiltonian actions.\cite{petruzielo2012semistochastic,wei2018reduced,sabzevari2018orbital,wu2025hybrid}

For fixed $x$, define $D_{\mathrm w}(x)=\sum_{y\in\mathcal C_{\mathrm w}(x)}|H_{xy}|$. The cumulative distribution proportional to $|H_{xy}|$ is divided into $N_{\mathrm{eloc}}$ equal-probability strata. If $y_j$ is selected from stratum $j$, the weak contribution is estimated as
\begin{equation}
\widehat E_{\mathrm L,w}(x)
=
\frac{D_{\mathrm w}(x)}{N_{\mathrm{eloc}}}
\sum_{j=1}^{N_{\mathrm{eloc}}}
\operatorname{sgn}(H_{xy_j})\frac{\psi(y_j)}{\psi(x)}.
\label{eq:si_weak_estimator}
\end{equation}
Repeated selected connections are combined before wave-function evaluation. The stratification is unbiased within the retained weak range, $\mathbb E[\widehat E_{\mathrm L,w}(x)|x]=E_{\mathrm L,w}(x)$, and the optimization estimator is
\begin{equation}
\widehat E_{\mathrm L}(x)
=H_{xx}+E_{\mathrm L,s}(x)+\widehat E_{\mathrm L,w}(x).
\label{eq:si_semistochastic_eloc}
\end{equation}
Its conditional expectation is $E_{\mathrm L}(x)-E_{\mathrm L,o}(x)$. Finite $N_{\mathrm{eloc}}$ contributes stochastic variance but no bias within the retained weak range, whereas $\varepsilon_2>0$ introduces deterministic truncation. The required wave-function evaluations are therefore restricted to the deterministic strong set and the sampled weak connections rather than the full $O(M^4)$ connected space. Setting both thresholds to zero evaluates the complete local energy deterministically.

\section{Computational Details}
\label{sec:computational_details}

\subsection{Electronic Hamiltonians and Initial Configurations}
\label{sec:si_electronic_models}

All calculations use the electronic Hamiltonian of the main text in an orthonormal spatial-orbital basis. Particle number and $M_S$ are fixed, and the projected backflow is evaluated in the corresponding highest-weight sector $S=M_S$. Hamiltonians supplied as FCIDUMP files are used in their published orbital ordering unless stated otherwise. The scalar term of each FCIDUMP is retained as given.

\paragraph{Water.}
The \ce{H2O} calculations use the all-electron $(10e,24o)$ cc-pVDZ Hamiltonians at symmetric O--H stretches $R/R_{\mathrm e}=1.0,1.5,2.0,2.5,3.0$. The geometries, orbital basis, and reference energies are those of Ref.~\cite{olsen1996full}; the FCIDUMP scalar term contains the nuclear repulsion energy. The initial configuration is the closed-shell occupation of the first five spatial orbitals in the supplied ordering. The reference orbital matrix used to initialize the backflow is the identity in this basis.

\paragraph{Hydrogen lattices.}
The hydrogen benchmarks contain 36 atoms at half filling in the STO-6G basis: a 36-site chain, a $6\times6$ square lattice, and a $3\times3\times4$ simple-cubic lattice, each with nearest-neighbor spacing $2.0\,\text{\AA}$. The one- and two-electron integrals are generated with PySCF.\cite{sun2018pyscf,sun2020pyscf} The atomic orbitals are symmetrically L\"owdin orthogonalized, the integrals are transformed to this orthonormal atomic-orbital basis, and matrix elements below $10^{-8}\,E_{\mathrm h}$ in magnitude are set to zero. The scalar Hamiltonian term is the nuclear repulsion energy. The initial configuration has one electron per site with opposite spins on the two sublattices; spatial orbital 0 is assigned to the spin-up sublattice.

\paragraph{Iron--sulfur clusters.}
The \ce{Fe2S2} and \ce{Fe4S4} calculations use the published $(30e,20o)$ and $(54e,36o)$ active-space Hamiltonians of Li and Chan in their original FCIDUMP ordering.\cite{li2017spinprojected} These are the same active-space models used in the recent high-accuracy Fe--S benchmarks of Ref.~\cite{zhai2026classical}. The FCIDUMP scalar terms are zero; following the convention of that work, the external core energies ($-4976.26532397\,E_{\mathrm h}$ for \ce{Fe2S2} and $-8105.56038966\,E_{\mathrm h}$ for \ce{Fe4S4}) are not included in the reported active-space energies. The initial configurations follow the broken-symmetry patterns associated with the corresponding benchmark states.

With zero-based spatial-orbital indices, the Fe $3d$ subsets are Fe1 $=2{:}6$ and Fe2 $=13{:}17$ for \ce{Fe2S2}; for \ce{Fe4S4} they are Fe1 $=2{:}6$, Fe2 $=7{:}11$, Fe3 $=24{:}28$, and Fe4 $=29{:}33$. For \ce{Fe2S2}, orbitals outside the two Fe subsets are doubly occupied in the initial configuration, with the five Fe1 orbitals occupied by spin-up electrons and the five Fe2 orbitals by spin-down electrons. For \ce{Fe4S4}, the initial configuration is the BS1 pattern $[\mathrm{Fe1}\uparrow,\mathrm{Fe2}\uparrow,\mathrm{Fe3}\downarrow,\mathrm{Fe4}\downarrow]$: orbitals outside the four Fe subsets are doubly occupied, Fe1 and Fe2 carry the predominantly spin-up occupations, and Fe3 and Fe4 the predominantly spin-down occupations. The additional spin-down electron occupies orbital 7 and the additional spin-up electron orbital 29.

\subsection{Numerical Parameters}
\label{sec:si_numerical_parameters}

The same sampling, wave-function, and optimization settings are used for all production calculations unless stated otherwise; the sampler and optimizer are not tuned by system.

\begin{table}[ht]
\centering
\caption{Common production parameters.}
\label{tab:si_parameters}
\small
\begin{tabular}{lll}
\toprule
Category & Quantity & Value \\
\midrule
Sampling & chains / evaluation samples & $4096/4096$ \\
 & initial thermalization & 4096 transitions \\
 & transitions per observation & 16 \\
 & initial $\alpha_0$ & 2.0 \\
 & $\alpha$ damping & 0.02 \\
 & evaluation probability $\beta$ & 0.5 \\
\addlinespace
Local energy & strong--weak threshold $\varepsilon_1$ & $10^{-3}\,E_{\mathrm h}$ \\
 & truncation threshold $\varepsilon_2$ & $10^{-12}\,E_{\mathrm h}$ \\
 & weak samples $N_{\mathrm{eloc}}$ & 1024 \\
\addlinespace
Wave function & hidden widths & $(256,256)$ \\
 & activation & SiLU \\
 & output-layer weights & normal, $\sigma=10^{-3}$ \\
\addlinespace
Optimization & SR shift $\lambda$ & $10^{-3}$ \\
 & predictor coefficient $\mu$ & 0.95 \\
 & learning rate $\eta$ & 0.05 \\
 & iterations & 5000 \\
\addlinespace
General & random seed & 0 \\
 & arithmetic & double precision \\
\bottomrule
\end{tabular}
\end{table}

\subsection{Reference Calculations}
\label{sec:si_reference_energies}

All reference energies correspond to the same Hamiltonians used in the VMC calculations. The \ce{H2O} FCI, UCCSDT, and MRCISD values are taken from Ref.~\cite{olsen1996full}. For the hydrogen lattices, we perform independent spin-adapted DMRG calculations with Block2\cite{zhai2023block2} using the singlet ($S=0$) SU(2)-adapted formulation.\cite{sharma2012spinadapted} Bond dimensions are $M=100,200,400,800,1600$, and each reported calculation uses a sweep-energy convergence threshold of $10^{-6}\,E_{\mathrm h}$. For \ce{Fe2S2}, the reference is the $M=12000$ SA-DMRG energy; for \ce{Fe4S4}, it is the extrapolated BS1 SA-DMRG energy from the corresponding active-space benchmark.\cite{li2017spinprojected,zhai2026classical}

\section{Post-optimization Analysis}
\label{sec:si_postanalysis}

\subsection{Born-Distribution MCMC Diagnostics}
\label{sec:si_born_mcmc}

To characterize Born-distribution autocorrelation of the optimized states, we freeze the final wave function and run 64 Metropolis--Hastings chains with invariant distribution proportional to $|\psi|^2$, initialized from saved production configurations. After 4096 thermalization transitions, each chain is propagated for 262144 transitions without thinning using a proposal kernel given by an equal mixture of single- and double-excitation moves. We estimate $\tau_{\mathrm{int}}$ for $2\log|\psi(x)|$ and report the corresponding MCMC effective-sample-size fraction.

\begin{table}[ht]
\centering
\caption{Born-distribution MCMC diagnostics for the optimized wave functions.}
\label{tab:si_born_mcmc}
\small
\begin{tabular}{lccc}
\toprule
System & Acceptance rate & $\tau_{\mathrm{int}}$ & $\mathrm{ESS}_{\mathrm{MCMC}}/N$ \\
\midrule
\ce{H2O}, $R/R_{\mathrm e}=1.0$ & 0.00598 & 2036 & $4.91\times10^{-4}$ \\
\ce{H2O}, $R/R_{\mathrm e}=1.5$ & 0.00769 & 1579 & $6.33\times10^{-4}$ \\
\ce{H2O}, $R/R_{\mathrm e}=2.0$ & 0.00859 & 1368 & $7.31\times10^{-4}$ \\
\ce{H2O}, $R/R_{\mathrm e}=2.5$ & 0.00686 & 1438 & $6.95\times10^{-4}$ \\
\ce{H2O}, $R/R_{\mathrm e}=3.0$ & 0.00399 & 1052 & $9.50\times10^{-4}$ \\
\ce{Fe2S2} & 0.08360 & 278 & $3.60\times10^{-3}$ \\
\bottomrule
\end{tabular}
\end{table}

The optimized \ce{H2O} wave functions exhibit substantially longer autocorrelation under Born sampling than \ce{Fe2S2}. Across the \ce{H2O} series, the acceptance rate and $\mathrm{ESS}_{\mathrm{MCMC}}/N$ vary nonmonotonically with bond length.

\subsection{Final Energy and Observables}
\label{sec:si_observables}

Final observables are evaluated from the optimized checkpoint by continuing the saved Markov chains with the final $\alpha$ and $\beta=0.5$; no additional thermalization is applied. The sampler retains the production strong-connection set defined by $\varepsilon_1=10^{-3}\,E_{\mathrm h}$, while the local-energy evaluator uses zero thresholds with $N_{\mathrm{eloc}}=0$. Sampling therefore uses the same retained connection set as during optimization, whereas the reported final energy uses the complete deterministic local energy. One evaluation configuration is collected from each of the 4096 saved chains.

The spin-resolved one-particle reduced density matrix is $\gamma_{pq}^{\sigma}=\langle a_{p\sigma}^{\dagger}a_{q\sigma}\rangle$, and the spin-summed matrix is $\gamma_{pq}=\sum_\sigma\gamma_{pq}^{\sigma}$. After Hermitization, $\gamma\leftarrow(\gamma+\gamma^\dagger)/2$, the natural occupations are its eigenvalues. Diagonal occupations and off-diagonal one-body contributions are evaluated with the same normalized importance weights used for the energy.

Orbital spin correlations are $C^S_{pq}=\langle\bm S_p\!\cdot\!\bm S_q\rangle$. Longitudinal terms are evaluated from occupations and transverse spin-exchange terms from local wave-function ratios; $\langle\hat S^2\rangle=\sum_{pq}C^S_{pq}$. For the hydrogen lattices, orbital pairs are grouped by Manhattan distance $d$ in the finite lattice and Fig.~4(b) of the main text reports $(-1)^d\overline{C^S}(d)$, averaged over distinct pairs at distance $d$. For the Fe--S clusters, site correlations are obtained by summing over the five Fe $3d$ orbitals on each site, $C^S_{IJ}=\sum_{p\in I}\sum_{q\in J}C^S_{pq}$.

\bibliography{reference}